# The Hitchhiker's Guide to Monoculture: AI Homogenizes Syntax, Not (Necessarily) Semantics

Gordon Burtch*

Questrom School of Business, Boston University, Boston, MA 02215

*To whom correspondence should be addressed. Email: gburtch@bu.edu

## Abstract

Large language models (LLMs) have been widely reported to homogenize human expression and thought. However, I argue that convergence in language need not imply convergence in ideas, and existing evidence rarely distinguishes between the two. I demonstrate this through a study of software development, where AI assistants have diffused fastest and where syntax can be readily separated from semantics (conceptual approach or intent). Using Kaggle contest submissions from 2019 to mid-2026, I first document convergence toward the random seed value 42, consistent with LLMs reinforcing a longstanding programming-culture convention associated with Douglas Adams' comedy novel *The Hitchhiker's Guide to the Galaxy*. I then measure homogenization in code syntax and approach more generally, quantifying within-contest code submission similarity using term-frequency–inverse-document-frequency (TF-IDF) n-gram representations, which capture surface syntax, and Voyage code-3 retrieval embeddings, which capture intent and approach. Submissions have become more alike in literal syntax, but I find no evidence of homogenization in approach or intent. I conclude that shared AI tools are homogenizing how Kaggle contestants express their solutions, not the problem-solving strategies they employ.

## Significance Statement

Large language models are increasingly believed to be homogenizing human expression and ideation. However, much of the evidence for this has focused on written text. Syntactic homogenization does not necessarily imply semantic homogenization: a single idea can be expressed with different words, while similar words can be used to express different ideas. Software code is a useful context to examine this distinction because neural embeddings enable separate measurement of code semantics and syntax. I study Kaggle, a platform for online coding contests, considering tens of thousands of code submissions spanning six years bracketing the emergence of AI. I find that the syntax of code submissions has become more similar, yet solution approaches remain diverse.

A rapidly growing literature finds that large language models (LLMs) are homogenizing human language and expression. Stories written with AI-generated ideas are more alike (1), ideas generated with LLM support are more similar across people (2), scientific writing has begun to incorporate LLMs' vocabulary (3), and variance in writing style has generally contracted across social media, news, and scientific text since ChatGPT's release (4).

These findings have been broadly interpreted as evidence that AI is narrowing the space of human ideas (5–7). However, that conclusion may be premature. Convergence in linguistic expression does not necessarily imply convergence in the ideas being expressed. An LLM may standardize vocabulary, phrasing, or grammatical structure while users continue to address different topics or articulate distinct ideas. Conversely, if two pieces of text employ similar wording, they may nonetheless make distinct or even opposite points. This observation underscores the importance of distinguishing between homogenization in surface form and homogenization in semantic content.

Prior work on LLM-driven homogenization has generally not sought to separate word choice from semantic content, perhaps in part because the distinction is difficult to operationalize in prose. With software code, however, that separation becomes more tractable. Code has literal syntax, e.g., variable names, and it also has more abstract semantics, namely its functional or conceptual content, i.e., what the code is intended to accomplish. Homogenization in the former is generally easy to measure. Further, homogenization in the latter is now increasingly straightforward to operationalize as well, leveraging modern code-retrieval embeddings. Understanding software code homogenization is also substantively important, in its own right, given developers' increasing reliance on AI coding assistants. With these points in mind, I examine LLM-driven homogenization of software code by separating surface syntax from semantics, using public submissions to Kaggle machine-learning competitions. In Kaggle competitions, participants work on the same well-defined task under common performance benchmarks. This setting makes it possible to compare the implementation and higher-level approach of completed human work while holding the problem relatively constant.

Concerns about code homogenization are particularly important because they connect to a long-standing discourse on technological monocultures. Reliance on shared software artifacts and algorithms creates two related ecosystem-level risks. Technological monoculture arises when many actors depend on the same component, creating a single point of failure, as recent disruptions involving Amazon Web Services and CrowdStrike illustrate (8). Algorithmic monoculture arises when decision-makers rely on the same algorithm or decision rule, causing errors and biases to become correlated; for example, a shared recruiting or lending model may reproduce the same distortion across employers or lenders (9–11). AI-assisted programming potentially links these risks. LLM assistants may standardize low-level implementation choices, including libraries, defaults, names, and code structure, and may also push developers toward shared higher-level solution strategies. Whether this is happening remains unclear, however, because users of AI assistants may prompt differently and adapt model outputs rather than adopting them verbatim (12). As such, the observation that LLMs respond homogeneously to a given prompt in a fixed context does not necessarily imply that working with AI assistants will result in homogeneous completed work. Distinguishing convergence in implementation from convergence in substantive approach is therefore essential.

Existing research has shown that LLMs are changing software development, but it has focused primarily on individual productivity (13–19), task allocation (20, 21), developer experience (22), generated-code quality and security (23, 24), as well as the decline of software knowledge communities (25). Further, while more recent literature examines generative monoculture, the tendency of language models to produce homogeneous outputs in response to prompts, e.g., Jiang et al. (26) find that repeated samples from the same model and outputs from different models converge on similar content, and Wu et al. (27) document related concerns in code generation, they only establish a plausible mechanism for AI-driven monoculture. They do not establish that these outputs make their way into developers' final code.

Large-scale field evidence for homogenization of human work product remains relatively limited, and the studies that have been conducted have not attempted to separate which dimensions of output are homogenizing. For example, Sourati et al. (4) document declining linguistic diversity at scale across social media, news, and scientific text. The closest large-scale analog to the present study is work by Zhou and Lee (28) and Zhou, Lee, and Gu (29), who examine the adoption of text-to-image generative AI on a large digital art platform. They show that generative AI can compress some dimensions of creative output while expanding others. However, those authors do not distinguish convergence in surface form from convergence in underlying ideas, and their open-ended setting makes changes in convergence harder to interpret. Artists generally do not work toward a common objective, and changes in platform conditions or audience demand may alter both what artists choose to create and the approaches they take when pursuing those shifting objectives. Experimental and observational studies of AI-assisted writing similarly document changes in textual diversity (30, 31). Taken together, this work does not establish whether apparent homogenization reflects convergence in linguistic expression, convergence in ideas, or both.

Whether LLMs' homogeneous output propagates into users' final work depends on how users respond to LLMs, e.g., accepting it verbatim versus iterating and modifying it. Jahani et al. (12) show that users adapt their prompts as model capabilities evolve, especially when tasks provide clear objectives and feedback. LLMs' homogeneous output may therefore expose users to similar initial suggestions but need not produce identical final choices. Users' adaptation may preserve variation.

Incentives also matter for how individuals will interact with LLMs. Gartenberg et al. (32), analyzing article submissions and peer review at a major academic journal around the emergence of LLMs, argue that academic "publish or perish" incentives may encourage greater output while reducing writing quality and narrowing topic span. Notably, Kaggle competitions combine some factors that may promote convergence, yet they also include factors that may counteract convergence. Participants share a narrow task, public examples, and visible performance benchmarks, making convergence easier to observe. At the same time, leaderboards and rapid feedback reward differentiation. This tension creates a clear possibility: AI assistance may standardize routine implementation choices without compressing the set of substantive solutions competitors pursue.

I test this possibility using 61,565 public Python submissions to 285 Kaggle competitions from 2019 through mid-2026. I first document rapid convergence on the arbitrary random-seed value 42 after ChatGPT's release, consistent with AI assistants reinforcing a shared programming convention. I then measure within-contest similarity in two spaces: TF-IDF token representations capture literal syntax, whereas code-retrieval embeddings capture apparent intent and approach. Submissions become more alike in syntax, but I find no comparable contraction in semantic diversity. The results show that generative monoculture can propagate into completed human work unevenly across levels of abstraction: common AI tools may standardize how solutions are expressed and assembled without necessarily narrowing the strategies users pursue.

## Results

**A case study of seed 42.** I begin with a simple but revealing case study: the value assigned to a random seed. Seeds make computational results reproducible by controlling stochastic operations such as initialization, resampling, and data splitting, yet developers generally regard the particular value as inconsequential. Recent work has shown that seed choice can affect the quality and stability of training and fine-tuning with transformer-based models (33, 34). Fig. 1 plots the share of Kaggle submissions that use 42 at least once. The value was already popular before widespread AI adoption because of its association with Douglas Adams' novel *The Hitchhiker's Guide to the Galaxy*, but its use accelerated sharply around ChatGPT's release. By May 2026, 91.9% of seeded submissions used 42. Its share rose by 0.31 percentage points per month; afterward, the monthly increase roughly tripled to 0.92 points ($P < 0.001$; Table 1). The monthly share likewise exceeded 70% in early 2026 in a broad sample of GitHub machine-learning repositories (SI Appendix, Fig. S1). The rapid convergence on this arbitrary convention provides a particularly clear case study of AI-associated homogenization. I next examine whether the pattern extends beyond a single implementation detail to broader features of machine-learning code.

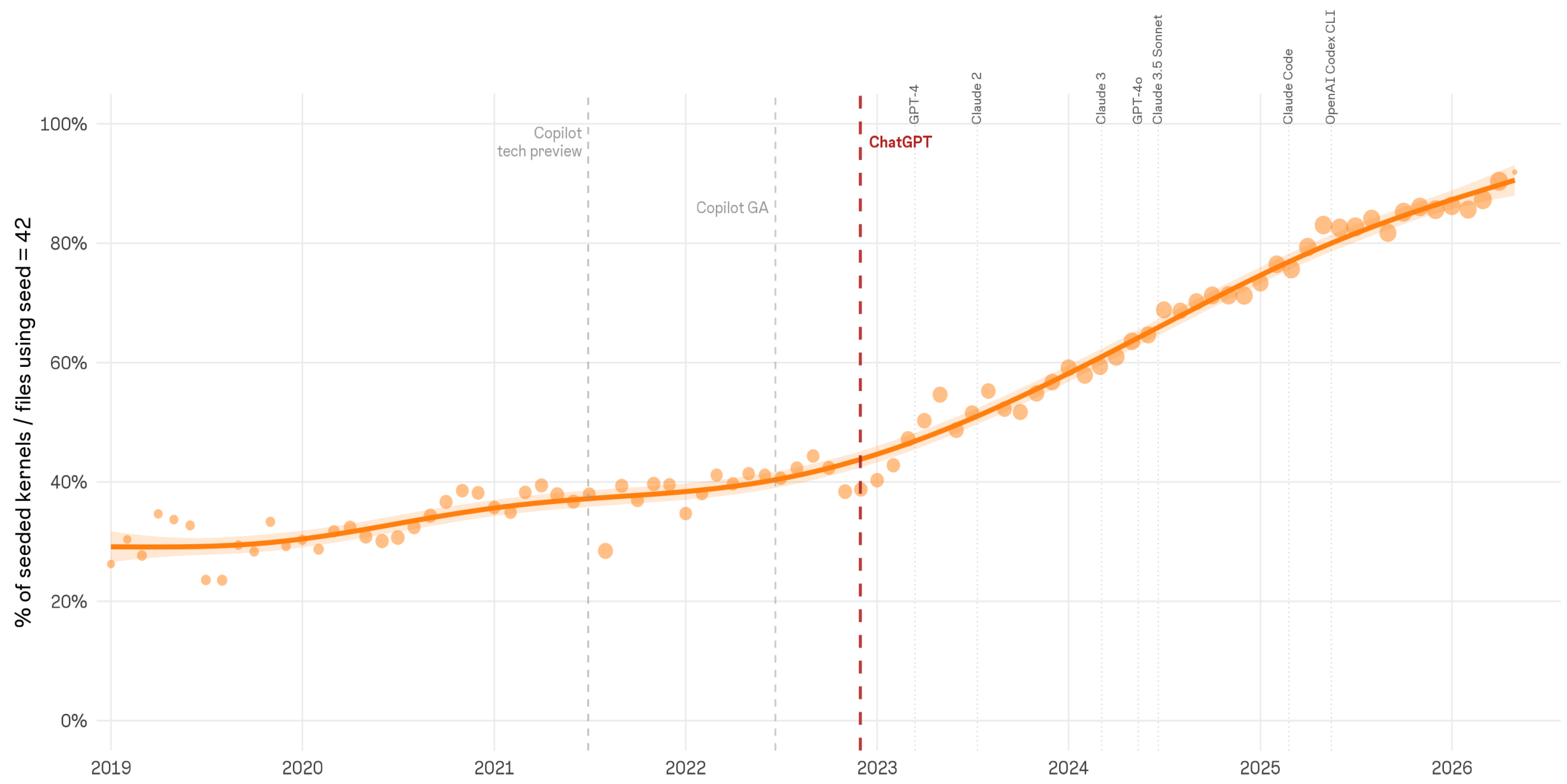

***Fig. 1. The percentage of seeded Kaggle kernels that employ seed = 42 at least once.*** *The share rose sharply after the introduction of AI coding assistants and reached 91.9% in May 2026.*

**Homogenization of syntax versus semantics.** To distinguish surface convergence from convergence in approach, I represent each submission in two ways. A TF-IDF n-gram representation captures literal syntax, whereas a Voyage code-3 embedding captures the apparent intent and approach of the code. I validate this distinction using synthetic pipelines that combine five modeling approaches with five naming conventions (SI Appendix, Fig. S3 and SI Appendix, Table S1). Voyage places pipelines closer together when they implement the same approach under different variable names; TF-IDF instead places them closer when they share a naming convention despite implementing different methods. Before constructing either representation, I mask seed values and remove comments, ensuring that neither seed results nor comment style mechanically determine measured similarity. I then calculate two within-contest outcomes: mean pairwise cosine distance, which measures average separation between submissions, and effective spectral rank (35), which measures the number of independent latent directions spanned by a contest's submissions.

After AI coding assistants became widely available, Kaggle submissions grew more similar in literal syntax but not in semantic content (Fig. 2). Their collective latent span shows the same separation: effective rank declines in syntactic space while remaining essentially unchanged in semantic space (Fig. 3). My interrupted time-series estimates indicate a post-release trend change of −0.0017 per month in TF-IDF pairwise distance ($P = 0.005$), with no significant change in Voyage space. TF-IDF effective rank similarly falls by 0.160 per month relative to its prior trend ($P < 0.001$), again without a significant change in Voyage (Table 1). I obtain the same effective-rank pattern after accounting for contest topics and objectives by residualizing each measure, through cross-fitted regression, against a contest-description embedding and structured contest characteristics. Taken together, these patterns describe convergence in how solutions are expressed and assembled, not in the substantive strategies competitors pursue. AI assistants appear to be standardizing the connective tissue of machine-learning work while leaving its central modeling decisions comparatively heterogeneous.

I also observe these results in annual kernel density plots of the two measures (SI Appendix, Fig. S2). A systematic shift toward pairwise similarity is clear in the TF-IDF space, beginning in 2023, yet the Voyage distribution remains stable; the contraction in contest-level effective rank occurs only in the TF-IDF space. These patterns suggest that, while AI may be inducing convergence in syntax, the solution space that contest participants explore remains largely undiminished.

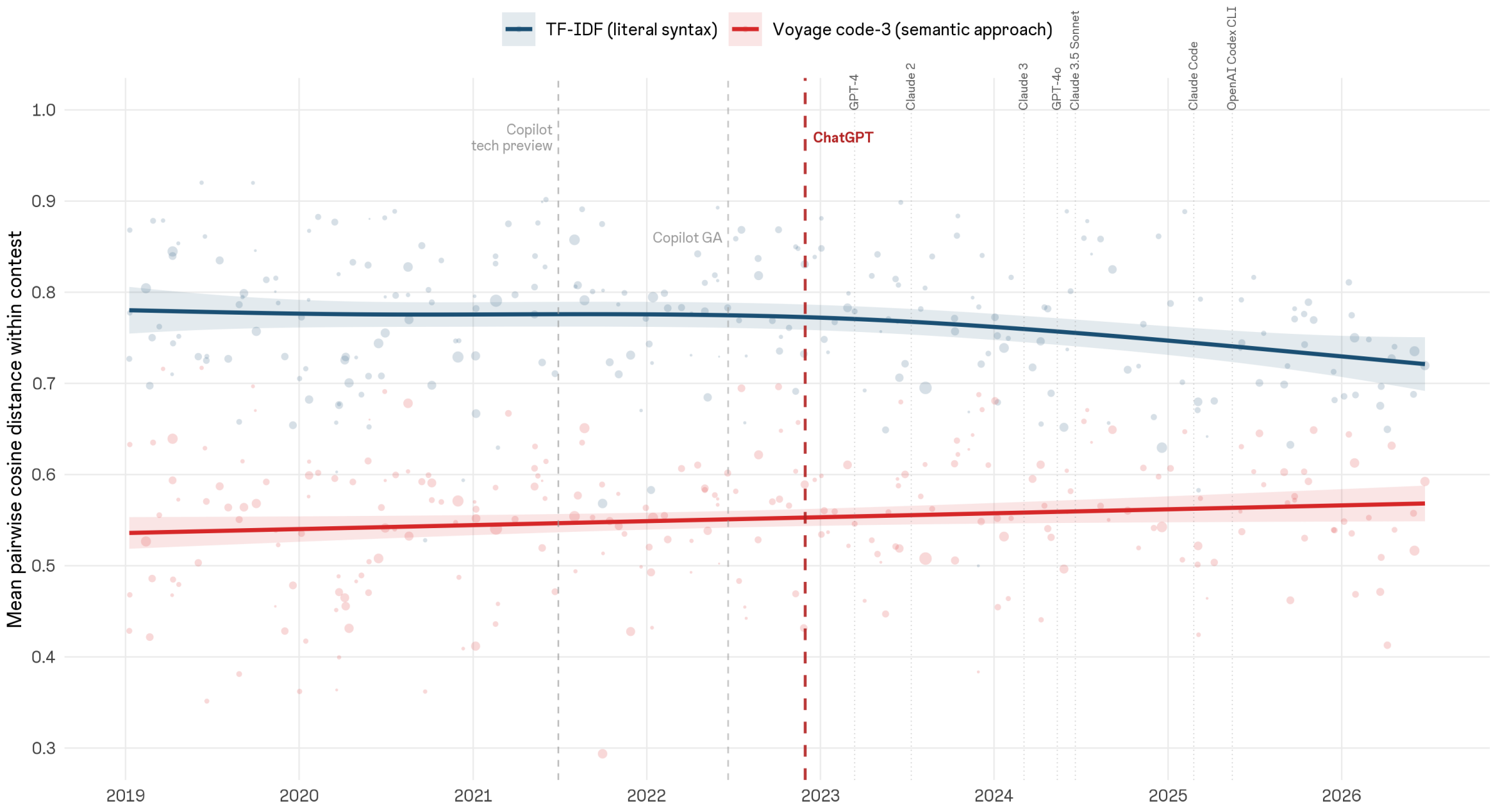

***Fig. 2. Within-contest mean pairwise cosine distance among Kaggle contest submissions over time.*** *Each scatter point is one contest; lower values indicate less pairwise novelty / more pairwise similarity among submissions. TF-IDF (syntactic) distance declines post-AI, while Voyage (semantic) distance remains unchanged.*

***Table 1. Interrupted time-series estimates.*** *Ordinary least squares regressions with a monthly time trend centered at the release of ChatGPT (30 November 2022), a post-release indicator, and their interaction. I report coefficients, 95% confidence intervals in square brackets, and a p-value in parentheses. All statistical tests are two-sided and based on heteroskedasticity-consistent (HC3) robust SEs. The seed-42 share is measured in percentage points; the embedding-based measures are per Kaggle competition. The Methods section under Statistical Tests provides the exact regression specification.* Significance markers: *** $P < 0.001$, ** $P < 0.01$, * $P < 0.05$, + $P < 0.10$.

| | **(1) Seed-42 share (pp)** | **(2) Pairwise dist., TF-IDF** | **(3) Pairwise dist., Voyage** | **(4) Eff. rank, TF-IDF** | **(5) Eff. rank, Voyage** |
|---|---|---|---|---|---|
| Intercept ($\beta_0$) | 42.364***<br>[40.757, 43.971]<br>(P < 0.001) | 0.7765***<br>[0.7542, 0.7989]<br>(P < 0.001) | 0.5556***<br>[0.5311, 0.5801]<br>(P < 0.001) | 33.468***<br>[32.352, 34.583]<br>(P < 0.001) | 25.182***<br>[24.289, 26.075]<br>(P < 0.001) |
| Time ($\beta_1$) | 0.315***<br>[0.247, 0.383]<br>(P < 0.001) | 0.0000<br>[−0.0008, 0.0009]<br>(P = 0.928) | 0.0005<br>[−0.0005, 0.0015]<br>(P = 0.313) | 0.050*<br>[0.004, 0.095]<br>(P = 0.034) | −0.004<br>[−0.035, 0.026]<br>(P = 0.783) |
| Post-ChatGPT ($\beta_2$) | 0.165<br>[−2.224, 2.554]<br>(P = 0.891) | 0.0099<br>[−0.0233, 0.0431]<br>(P = 0.557) | 0.0105<br>[−0.0229, 0.0439]<br>(P = 0.537) | −0.759<br>[−2.657, 1.138]<br>(P = 0.431) | −0.999<br>[−2.485, 0.486]<br>(P = 0.186) |
| Time × Post ($\beta_3$) | 0.918***<br>[0.819, 1.017]<br>(P < 0.001) | −0.0017**<br>[−0.0029, −0.0005]<br>(P = 0.005) | −0.0007<br>[−0.0021, 0.0006]<br>(P = 0.284) | −0.160***<br>[−0.239, −0.081]<br>(P < 0.001) | 0.044<br>[−0.019, 0.106]<br>(P = 0.170) |
| Observations | 89 months | 252 competitions | 252 competitions | 205 competitions | 205 competitions |

**What aspects of syntax converge?** I next decompose the syntactic homogenization in two ways (SI Appendix, Fig. S6 and SI Appendix, Table S4). First, I explore homogenization across the outcome distribution: it is strongest at the 10th percentile of homogeneity and weakest at the $90^{th}$ percentile. This suggests that the most syntactically distinctive submissions remain equally distinctive post-AI. Second, across different syntactic components, I find that homogenization primarily arises in relatively mundane code elements. Effects are largest for variable names, data loading, manipulation, and plotting functions, and keyword-argument names. However, function calls involving machine-learning algorithms are notably *not* affected. Thus, homogenization appears greatest in the aspects of code that a contestant likely does not specify in detail when prompting an AI coding assistant to implement a solution.

**Homogenization of syntax is not driven by who competes.** One factor that could contribute to a decline in average pairwise novelty would be the arrival of many new, AI-reliant (vibe coding) Kaggle participants. However, restricting the sample to veterans, such that every contestant has at least one submission predating ChatGPT's release, and applying contestant fixed effects, I observe consistent results (Fig. 4): average TF-IDF distance declines by 0.032 by the first half of 2026 ($P = 0.04$), while Voyage distance remains unchanged ($-0.002$, $P = 0.92$).

**Linking homogenization to AI usage.** My evidence that AI drives homogenization has thus far been associative, as I show that the trend shifts began around the time AI started to expand. To measure AI's impact more directly, I train a classifier that detects LLM-generated code comments (SI Appendix, Figs. S4–S5 and Table S2). Conditional on the contestant and competition, a one-unit (full-scale) increase in a submission's AI-comment score is associated with a 12.5-percentage-point increase in the probability that the submission sets a seed value to 42 ($P = 0.001$; SI Appendix, Table S2). Aggregating to the contest level, a ten-percentage-point increase in AI share is associated with a 0.017 reduction in mean TF-IDF pairwise distance ($P < 0.001$) and a 0.7-point decline in TF-IDF effective rank ($P = 0.009$). The Voyage associations are much weaker; although the decline in pairwise distance in Voyage space is significant, it is still only half the size of the TF-IDF effect (a 0.009 reduction, $P = 0.028$), and I again observe no statistically significant change in effective rank ($P = 0.147$) (SI Appendix, Fig. S5).

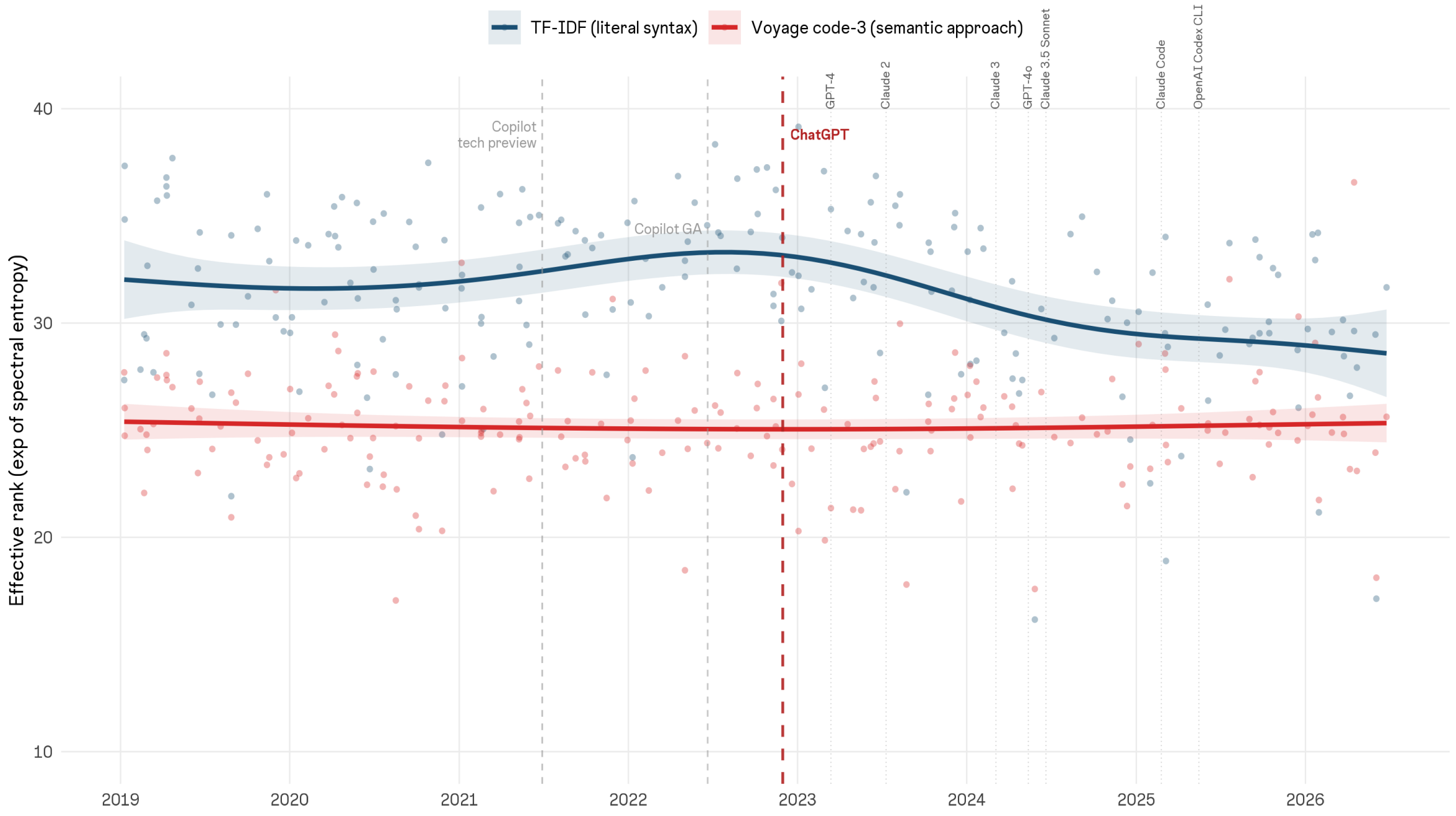

***Fig. 3. Per-contest effective spectral rank among Kaggle contest submissions over time.*** *TF-IDF effective rank falls markedly post-AI; Voyage effective rank remains flat.*

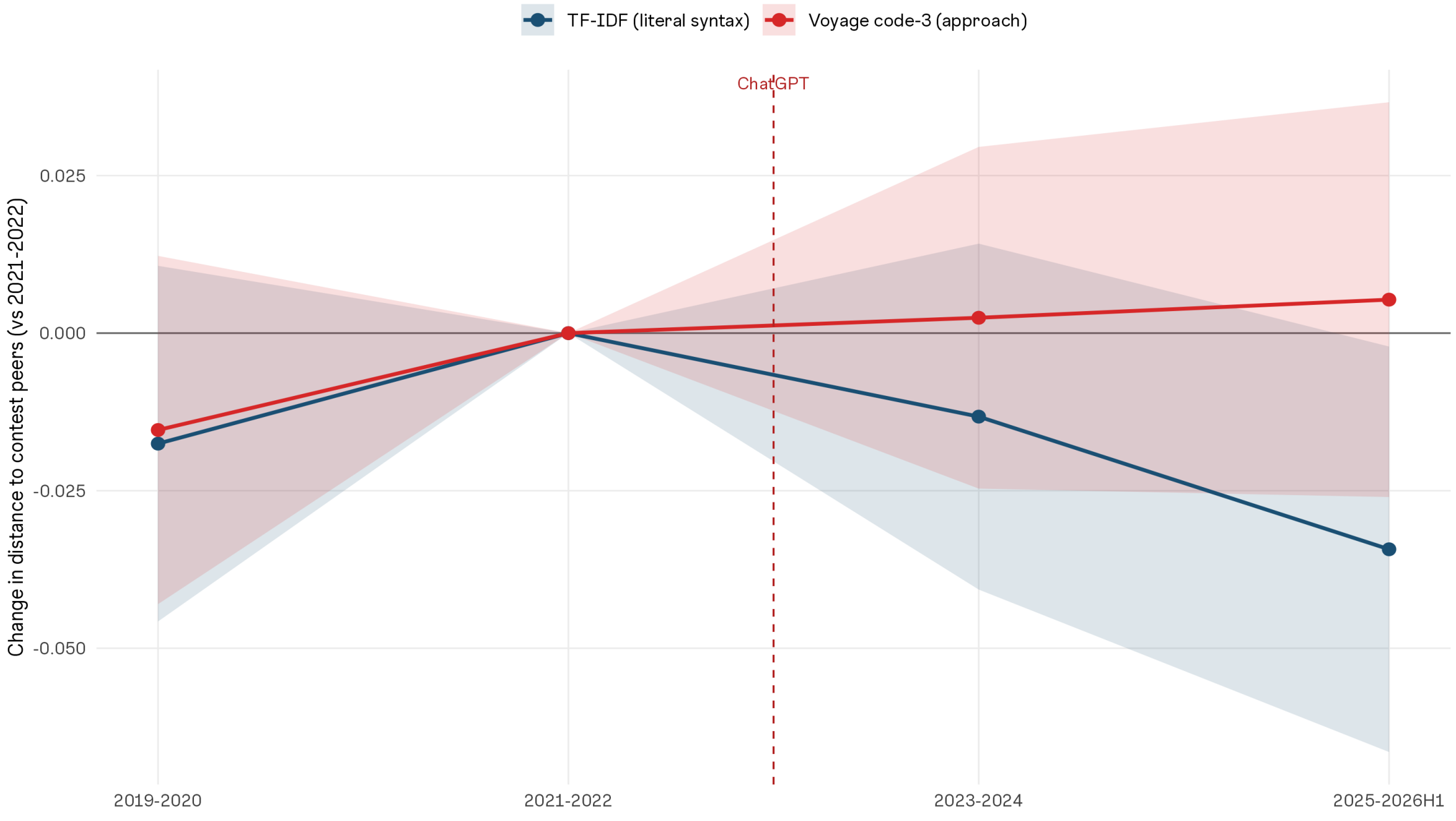

***Fig. 4. Biennial (two-year) change in within-contestant distance to contest peers among veteran contestants.*** *Pre-ChatGPT participants only, with contestant fixed effects. TF-IDF distance declines; Voyage distance remains flat. Bands are 95% confidence intervals, clustered by contest. I use two-year periods here to address data sparsity.*

## Discussion

AI coding assistants appear to be standardizing code style and implementation while leaving higher-level solution approaches diverse. The distinction matters beyond software: my findings suggest that reports of AI homogenizing human expression and thought (4–6) should be read with caution.

Prior work on AI-induced homogenization often treats surface similarity as evidence of conceptual convergence (2, 27, 30). The distinction I draw reconciles the dramatic spread of seed 42 with the stability of semantic diversity. Both reflect coordination, but at different levels of abstraction: a shared arbitrary default shows that developers draw from a common source of recommendations, not that they solve problems in the same way.

I argue more broadly that LLM-induced homogenization can take different forms. AI coding assistance may coordinate developers around common defaults, templates, libraries, and implementation patterns while having little influence on the overarching strategies they use to solve problems. Conversely, conceptual convergence may increase even when surface-level implementations appear diverse. Distinguishing these layers is necessary for understanding the ecosystem-level consequences of generative AI: the relevant question is not simply whether AI makes work products more similar, but where and how that similarity arises.

I interpret the pattern of results as reflecting how developers specify versus delegate choices in their prompts to LLMs. A developer may specify an algorithm while leaving the random seed or data-loading routine unstated. Choices developers retain remain diverse; choices delegated to the model converge on common defaults. Recent experimental findings that ChatGPT polishes students' solutions without narrowing their ideas (36) support a similar conclusion: prose homogenizes because it is delegated, whereas high-level ideas do not homogenize because they are not delegated. AI adoption can therefore standardize technological conventions while algorithmic strategies remain varied. Importantly, the boundary between what users specify and what they leave to the model will vary across contexts and over time. A developer who cares that a bug is fixed, but not how, may leave the method unspecified. Users may also defer to an LLM's choice of approach when they believe it knows better. In such cases, delegation extends to the problem-solving strategy itself, making approach homogenization more likely.

My findings also imply a broader governance lesson: ecosystem-level risk can arise even when each individual LLM suggestion appears harmless in isolation. System audits typically examine whether an AI recommendation is correct, secure, maintainable, or legally usable for its recipient. Worries over monoculture add another question: what happens when many people adopt the same individually acceptable recommendation? Evaluating AI-assisted development at the ecosystem level requires attention not only to individual output quality, but also to the dependence those outputs create across users, projects, and organizations.

At the same time, my results suggest that AI-induced monoculture very likely depends on context. On the one hand, Kaggle contests have a mix of factors that should promote homogenization. However, Kaggle also provides strong incentives for contestants to differentiate themselves. Those incentives may help preserve conceptual diversity, even as implementation details standardize. The point is that AI coding assistance may homogenize software unevenly across settings: in routine production environments and low-incentive coding tasks, conceptual convergence may be more pronounced if developers lack any incentive to be unique or creative; in competitive or heavily peer-evaluated environments, developers may remain active participants, using AI to accelerate implementation without fully surrendering the exploration of alternative solution strategies.

One possible ecosystem-level consequence is collective overfitting to publicly visible data. In the SI Appendix, Fig. S7 and Table S5, I show that competitions with more homogeneous submissions experience larger collective declines from the public to the private leaderboard. Because this evidence is cross-sectional, I interpret it as an association rather than a causal effect. Even so, the result illustrates why monoculture can matter beyond the quality of any individual artifact. Uniformity may remain unobjectionable under ordinary conditions while amplifying exposure to a shared blind spot.

Several limitations are worth noting. First, my Kaggle panel reflects voluntary publication; contestants choose whether to publish their kernels and which versions to publish. I address this in part by restricting my analysis to original (non-fork) submissions and to non-Playground competitions, but selection on publication remains a concern. That said, I examine publication rates over the study period (SI Appendix, Fig. S9) and find that they have not shifted over time in a manner that would suggest this selection drives my results. Second, the two embedding spaces I employ, TF-IDF n-grams and Voyage code-3 retrieval embeddings, are reasonable choices, but alternatives exist; embeddings derived from abstract syntax trees, execution traces, or trained on functional equivalence might produce different results. Notably, though, I validate the efficacy of my measures for recovering different aspects of code similarity and evaluate two alternatives that yield similar findings (SI Appendix, Fig. S3 and Table S3).

## Materials and Methods

**Kaggle Contest Submissions.** I obtained Kaggle kernel source code and metadata from the public Meta-Kaggle and Meta-Kaggle-Code data dumps in early July of 2026. My primary sample comprises publicly posted Python kernels submitted to Kaggle competitions. I link each kernel to a contest via the Kernel Version Competition Sources table. I then restrict the sample to competitions in Kaggle's Featured, Research, Recruitment, and Analytics categories, excluding Playground (prize-free) and 'getting-started tutorial' competitions, given they lack the incentive structure of genuine competitions. I also exclude essay and report competitions, where submissions are primarily prose rather than code (e.g., the 2023 Kaggle AI Report). To mitigate concerns that pairwise similarity may simply reflect copying via Kaggle's forking mechanism, I also exclude explicitly forked kernels and near-duplicate kernels (SI Appendix, Fig. S8). My resulting panel comprises 61,565 unique contest submissions to 285 distinct competitions, from 28,926 unique users. Among these submissions, 25,300 (41.1%) explicitly set a random seed. Of those, 62% (15,737) set the literal value 42 at least once, with the vast majority of these occurring in recent months and years.

**Code Embeddings.** For the Kaggle contest panel, I construct two embeddings for each submission that capture, respectively, surface-level syntactic and higher-level semantic similarity. Before embedding, I again strip comments and markdown from every submission so both embedding measures capture convergence in the code alone (note that I use comment text only in one set of analyses, where I train an LLM-comment classifier, described in SI Appendix, Extended Methods). When constructing these embeddings, I replace every detected random-seed value with a common placeholder token before constructing either similarity representation. This ensures that the sharp rise of the literal value 42 documented above is distinct from my measured convergence in either vector representation. For surface-level syntactic embeddings, I construct a sparse TF-IDF representation of each kernel over unigrams and bigrams of code tokens, following standard practice (37). I apply sublinear term-frequency scaling, inverse-document-frequency weighting, minimum and maximum document-frequency thresholds (a minimum of 10 documents, and a maximum of 95% of documents), and retain the top 20,000 scaled features. I then reduce the resulting sparse matrix to 1,024 dimensions via truncated singular value decomposition (38).

For higher-level semantic embeddings, I obtain embeddings of the same contest submissions from Voyage AI's voyage-code-3 model (39) via Voyage's paid API. These are 1,024-dimensional embeddings derived from a pretrained embedding model designed specifically for code retrieval tasks. I embed each contest submission as a single vector and, to accommodate the model's 32,000-token input limit, truncate submissions to the first 96,000 characters of code (roughly 32,000 tokens, conservatively). Note, however, that fewer than 2% of submissions exceed this threshold. Voyage-code-3 is trained with Matryoshka representation learning (40), which nests smaller embeddings within larger ones so that the leading dimensions carry the most information, loosely analogous to the variance ordering imposed by the truncated SVD on the TF-IDF space; this makes the two representations structurally parallel. Unlike TF-IDF, Voyage-code-3 is trained on a large code corpus with a retrieval objective, so it places submissions near one another in the embedding space when their code accomplishes more similar coding tasks, regardless of raw syntax. I obtain the same syntactic and semantic patterns when I substitute character n-gram TF-IDF and Cohere Embed v4 for the two focal spaces (SI Appendix, Table S3).

**Measures of Code Homogenization.** For each embedding type, I construct two complementary measures of within-contest homogenization.

*Submission-level pairwise distance.* For each kernel, *i*, submitted to contest, *c*, I compute novelty as the mean cosine distance between *i* and all other kernels submitted to the same contest. Because learned embedding models, particularly Voyage-code-3 in this case, are known to exhibit anisotropic embedding geometry, i.e., by default, the raw vectors share a strong common-axis component that mechanically inflates raw pairwise cosine similarity, the mean-centering I apply to all embedding measures removes the global common-mode component, which is the dominant source of this inflation. I then re-normalize all embeddings before computing cosine distances. This is standard practice in the embedding analysis literature (41, 42). Additionally, to avoid noise in the resulting measure, I enforce a minimum threshold of 20 public submissions per contest; that is, I exclude contests with fewer than 20 public submission kernels from the analysis.

Finally, I average pairwise distances within each contest. This within-contest aggregation ensures the resulting measures reflect only within-contest code similarity, not cross-contest similarity. Were I to construct pairwise comparisons *across* contests, e.g., at a point in time, my measure would be influenced by variation over time in the similarity versus novelty of cross-contest objectives. My approach ensures this is not the case here. A decline in mean within-contest pairwise distance over time indicates that submissions are becoming less distinctive relative to their contemporaneous peers tackling the same problem.

*Contest-level effective rank.* Adapting Roy and Vetterli's entropy-based rank (35) to the covariance spectrum, I define the effective rank of a contest's kernel embedding set as $\exp(H)$, where $H = -\Sigma_k p_k \log p_k$ and $p_k = s_k^2 / \Sigma_j s_j^2$. Here, $s_k$ are the singular values of the contest-mean-centered embedding matrix, so $p_k$ is the share of variance along the k-th principal direction. Note that $s_k^2$ is proportional to the corresponding covariance eigenvalue. This matrix is formed from the original embedding vectors, not the corpus-centered, unit-normalized vectors used for the pairwise cosine measure: the observation-wise normalization is deliberately omitted so that effective rank reflects the full spread of the embeddings rather than their angular similarity alone. Effective rank captures the number of independent directions along which kernels spread within the contest: lower values indicate that variation concentrates along fewer directions; higher values indicate that variation spans many directions. Effective rank is a within-contest measure that complements pairwise distance because contest submissions can be moderately close to one another, on average, yet still collectively vary along many independent directions in the embedding space, or vice versa. As such, the average pairwise distance measure and the effective rank measure jointly capture what can roughly be thought of as the 'volume' of code that a contest spans.

Because effective rank increases with sample size (a larger number of kernel embeddings can increase the likelihood that the embeddings span more orthogonal components merely by chance), I randomly sample 50 submissions per contest and use the resulting embeddings to calculate the rank measure. I repeat this process 3 times, resampling 50 kernels per contest without replacement, and ultimately report the mean result per contest. Additionally, to ensure a fixed embedding count per contest sample, I exclude contests with fewer than 50 public kernel submissions from this analysis.

**Statistical Tests.** For each monthly or contest-level series depicted graphically, I estimate an interrupted time-series (ITS) regression of the outcome, modeling the dependent variable as a function of a linear time trend, $t$ (in months, centered at ChatGPT's public release on November 30, 2022), an indicator for the post-release period, and the interaction between the two, employing heteroskedasticity-robust standard errors. Specifically, I estimate $y_t = \beta_0 + \beta_1 t + \beta_2 Post_t + \beta_3 (t \times Post_t) + \varepsilon_t$, with $Post_t$ being an indicator equal to 1 when $t \geq 0$ and zero otherwise. Under this parameterization, $\beta_1$ captures the pre-release trend, $\beta_2$ captures any discrete level shift at the time of release, and $\beta_3$ captures the change in trend thereafter. Because $t$ is centered on the ChatGPT release date, the intercept, $\beta_0$, reflects the counterfactual outcome implied by the pre-release trend at the release date; the fitted level at the release date itself is $\beta_0 + \beta_2$. Table 1 in the main text reports ITS estimates for the main outcomes. I also apply a more data-driven approach: binary segmentation breakpoint detection (43), allowing up to three mean-shift breaks under an asymptotic penalty at the 5% level. Finally, to understand the degree to which any homogenization may be driven by AI-enabled entry of new, novice contestants, I analyze pre-ChatGPT Kaggle 'veterans' to identify within-contestant code homogenization pre- versus post-AI. There, I regress each contestant's submission mean distance to the other submissions in the same contest onto a vector of dummies capturing two-year periods (to ensure sufficient support per indicator) and contestant fixed effects, with standard errors clustered by competition. Unless noted otherwise, the fitted trend curves displayed in all figures are penalized spline, generalized additive model (GAM) smooths (44).

## Data, Materials, and Software Availability

This study uses publicly available data released by Kaggle: the Meta-Kaggle dataset (45) and Meta-Kaggle-Code dataset (46), retrieved on 2 July 2026. I provide derived analysis inputs and all code in a public Code Ocean capsule (47).

## Author contributions

Author contributions: G.B. designed research, performed research, analyzed data, and wrote the paper.

## Competing interest statement

The author declares no competing interests.

## Acknowledgments

This research received no specific grant from any funding agency in the public, commercial, or not-for-profit sectors.

## Ethics

My study relies exclusively on publicly available data, software artifacts, and their associated metadata from Kaggle and GitHub. I do not intervene or interact with human participants, and my study does not constitute human-subjects research.

# Supporting Information Appendix for: The Hitchhiker's Guide to Monoculture: AI Homogenizes Syntax, Not (Necessarily) Semantics

Gordon Burtch*

Questrom School of Business, Boston University, Boston, MA 02215

*To whom correspondence should be addressed. Email: gburtch@bu.edu

**This PDF file includes:**

Extended Methods

Figs. S1 to S9

Tables S1 to S5

## Extended Methods

**Within-contest sample.** For the seed-choice analysis, I consider the full set of public Kaggle kernel submissions, timestamped by submission date. For more general measures of code homogenization, namely, within-contest similarity and contest-level code span measures, I filter the sample in three ways. First, I retain only completed competitions as of the data-collection date (July 2, 2026). I remove in-progress competitions primarily because a constructed measure of contest-level code span may be premature and mechanically lower than for completed contests. Second, within each retained competition, I retain only kernels created on or before the contest deadline while the competition was active. Roughly one quarter of contest-linked kernels are published *after* the contest deadline. For long-lived, popular competitions, post-deadline submissions accumulate over many years, so many submissions published in 2026 remain associated with competitions that closed as early as 2015; including these post-deadline submissions complicates timestamps that inform the construction of contest-level measures of code span. Finally, because a single participant may publish many distinct, incremental submissions associated with a single contest, I take the participant, rather than the submission, as the unit of analysis, representing each contestant by their most recent pre-deadline submission to the competition. I timestamp each retained competition by its submission deadline. With these restrictions, I obtain a final sample of 252 competitions with at least 20 participants to construct within-contest pairwise-distance measures, and 205 competitions with at least 50 participants to construct the contest span (spectral entropy) measure, over the 2019 to 2026 window.

**GitHub ML Repositories.** For the seed-choice analysis on GitHub, I create a panel of public Python files from repositories that use standard machine-learning or scientific-computing libraries (i.e., that import any of numpy, scikit-learn, tensorflow, pytorch, keras, jax, xgboost, lightgbm, catboost, or pandas). I sampled repositories through the GitHub Search API, stratified by buckets of repository star counts and time, from January 2019 through mid-2026. I cloned the repositories locally and parsed the code to identify explicit seed-setting calls. Each repository file is dated based on its repository's month of creation. Because files may be added to a repository after it was created, these date stamps are noisier than the submission timestamps I use in the Kaggle sample. My resulting panel includes 19,938 distinct repositories with 303,101 Python files. Approximately 2.7% (8,303) of these files set an explicit seed value. Of those, 2,485 set a seed of 42, with the majority concentrated in recent months and years.

**Random Seed Extraction.** I extract explicit seed-setting calls from each Python source file via regular-expression matching of invocation patterns for common machine-learning libraries, including random.seed(N) and seed = N kwargs for Python's built-in random module, np.random.seed(N) for NumPy, random_state=N for scikit-learn, tf.random.set_seed(N) and tf.set_random_seed(N) for TensorFlow, torch.manual_seed(N) and torch.cuda.manual_seed(N) for PyTorch. The pattern extractor also identifies NumPy Generator seeds (np.random.default_rng(N)), everything(N) and set_seed(N) helper wrappers (including PyTorch Lightning's), as well as capitalized seed constants (e.g., RANDOM_SEED = N or RANDOM_STATE = N), as long as they are defined outside a comment. Before applying the regex-based matching, I strip Python comments and triple-quoted docstrings so seed values are captured only if they appear outside a comment block. A file is thus marked as using "seed = 42" if any pattern captures the literal integer 42 anywhere in its uncommented source at least once. When creating the aggregate share figures, I compute the share of seeded files (or kernels) using each captured seed value in each period.

In Fig. S3, I validate that these two representations capture the aspects of code I have described: syntax (literal code, as written) versus semantics (the code's intent, i.e., what it does). To validate this, I construct realistic Kaggle-style machine-learning pipelines and re-embed them in each embedding space, applying a series of controlled edits in two dimensions: the machine-learning method implemented and the variable-naming convention used, while holding all other aspects of the surrounding pipeline fixed. In performing this analysis, I also ensure that variable naming conventions do not convey any information about the machine-learning method; I keep the names purposefully ambiguous. Considering the cross-product of five methods and five naming conventions, I show that two code samples are closer in Voyage embedding space when they share a modeling approach but use different variable-naming schemes. Conversely, code is closer in the TF-IDF representation space when two samples share a variable-naming scheme, even when they employ different machine learning methods.

**AI-comment classifier.** To measure AI usage directly, I train a classifier that detects LLM-generated code comments. I collect a random sample of Kaggle kernels submitted before ChatGPT's release in November 2022, where code comments are highly unlikely to have been LLM-generated. This set serves as a "purely human" sample spanning 248 contests. For the same contests, I prompt four LLMs (Claude Haiku 4.5, Claude Sonnet 4.6, Gemini 2.5 Flash, and GPT-4o) to generate one-shot solutions, providing each model with the contest title, overview, and dataset description. I extract all code comments from the human- and LLM-generated solutions, obtain character-level TF-IDF n-gram representations (3-, 4-, and 5-grams), and train a logistic-regression classifier of LLM-versus-human authorship. Five-fold cross-validation, with folds sampled at the contest level so that no contest contributes to both training and evaluation folds, yields a mean AUROC of 0.996. On ten held-out pre-2022 competitions, human kernels average a score of 0.14 (just 1.7% receive a score exceeding 0.5), whereas LLM-generated solutions average 0.78–0.83 (87–96% exceed 0.5). Because humans working with assistants will not always adopt LLM comments verbatim, I treat the classifier score as a lower-bound proxy for AI usage and label a submission AI-associated when its score exceeds 0.50, prioritizing recall over precision.

**Contest-description embeddings and cross-fitting.** To ensure that measured convergence is not an artifact of shifting contest topics or objectives, I residualize the contest-level outcomes with respect to a contest-description embedding and structured contest features. I embed each contest's title and description with Voyage's general-purpose text model (voyage-3) and retain its leading principal components as a proxy for the contest objective; structured features include the log number of teams, log prize, contest year, and host type. I use cross-fitted regression for residualization, estimating the residualizing model on held-out folds and applying it out of sample to avoid overfitting. Conditioning on these controls, the post-release trend change in TF-IDF effective rank remains −0.13 ($P = 0.01$), while the Voyage change is small and insignificant ($\beta = +0.004$, $P = 0.91$), as noted in the main text.

**Code-component decomposition.** To identify which parts of submission syntax homogenize, I associate each code token with one syntactic category using the abstract syntax tree: variable names, function calls, library (import) names, keyword-argument names, numeric argument values, and string argument values. I then group written keywords, operators, structural literals, and any bigram that can be equally associated with two or more categories into an 'unattributed' category. For each category, I re-weight its tokens using the same fixed inverse-document-frequency weights based on the original corpus, then project them through the fixed truncated-SVD basis of the main TF-IDF measure, so each component is measured over time in one consistent representation. Within function calls, I further separate those that reference a machine-learning algorithm, architecture, or optimizer from all others (data loading, manipulation, plotting, and utilities) to distinguish those related primarily to 'approach' from those more likely related to 'plumbing', e.g., data loading. The observed convergence in function calls arises entirely from the latter (non-ML calls homogenize significantly, $P < 0.001$, whereas ML-algorithm calls do not, $P = 0.43$).

**Contest-level AI-usage association.** To estimate the contest-level association between AI usage and convergence (Fig. S5), I relate each contest's share of AI-associated kernels (classifier score $\geq 0.50$) to its within-contest pairwise distance and effective rank. I residualize each representation with respect to the other, the contest-description embedding, and structured contest features (objective, size, reward, and host type), and report outlier-robust partial correlations. Throughout these secondary analyses, I retain only contests with the minimum submission counts stated above (20 for pairwise distance, 50 for effective rank), and contest-level regressions use one observation per contest.

# SI Figures and Tables

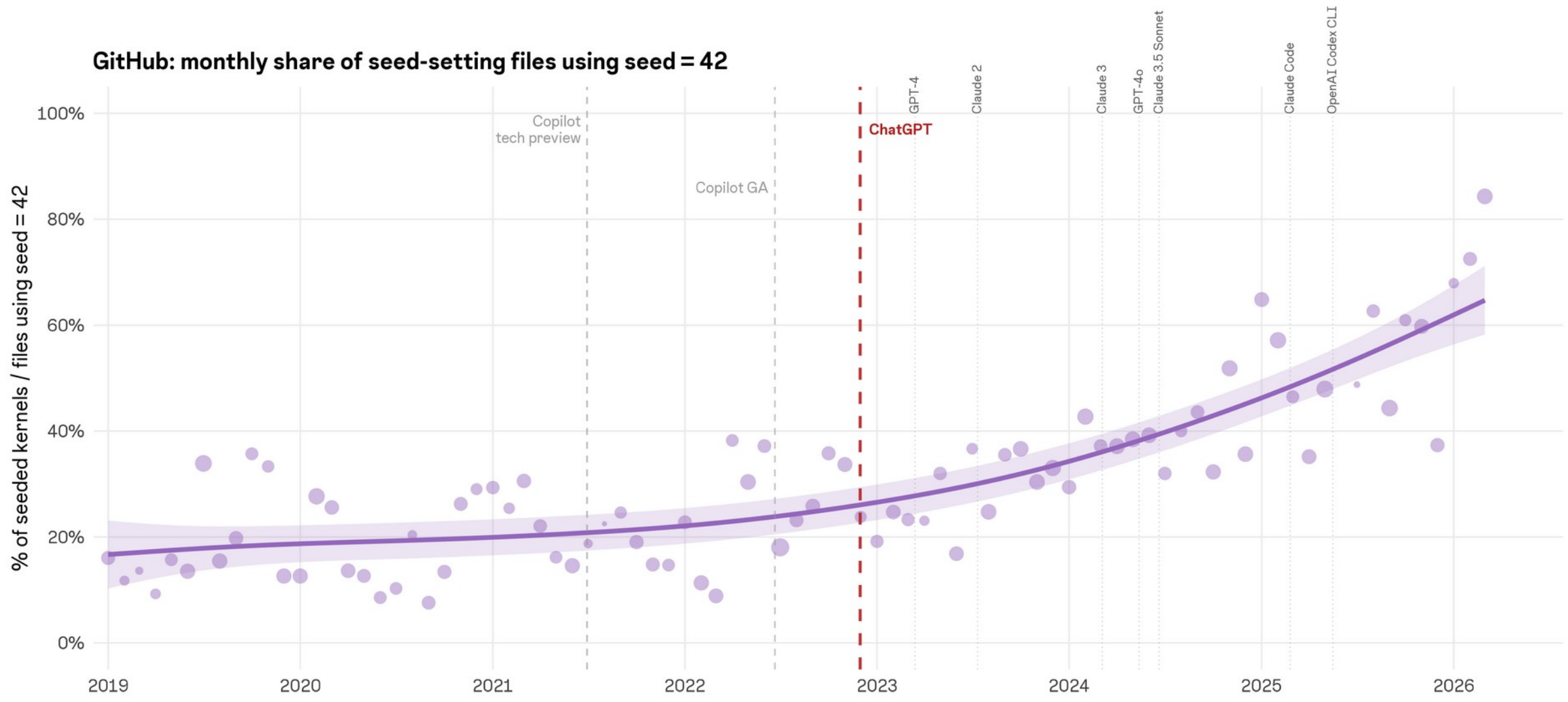

***Fig. S1. Share of seeded GitHub machine-learning repository files using seed = 42 at least once over time.*** *Each point is one month ($n = 86$ months). The share rises following the introduction of AI coding assistants and exceeds 70% in early 2026. An interrupted time series gives a pre-release trend of +0.21 percentage points per month ($P = 0.025$) and a post-release trend change of +0.89 ($P < 0.001$), the same pattern observed at Kaggle.*

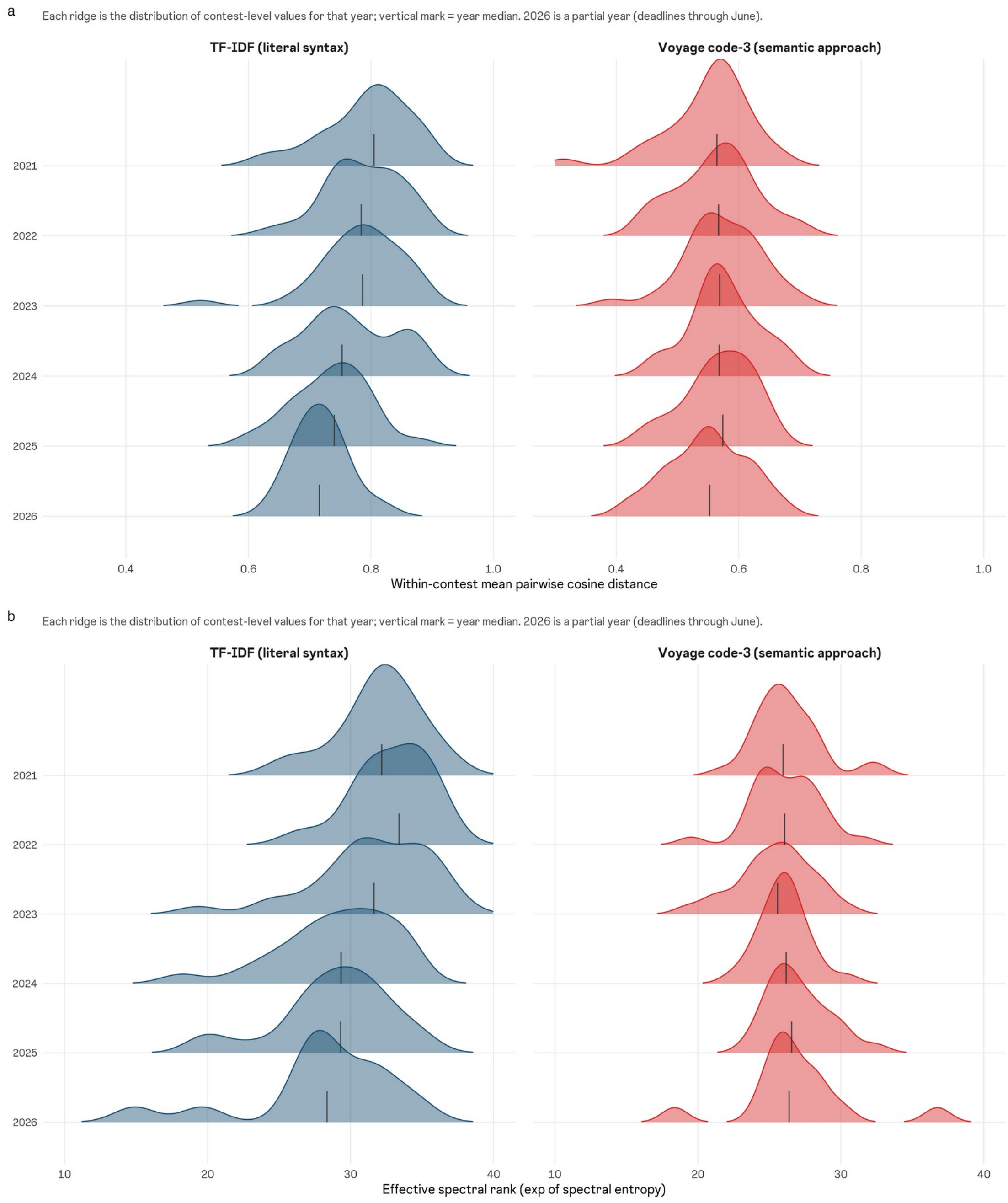

***Fig. S2. Annual densities of within-contest homogeneity. a***, *Mean pairwise cosine distance (n = 252 competitions).* ***b***, *Effective spectral rank (n = 205 competitions). Distributions are shown separately for each year and for both embedding spaces. Syntactic distributions shift after 2023; the semantic distributions are stationary.*

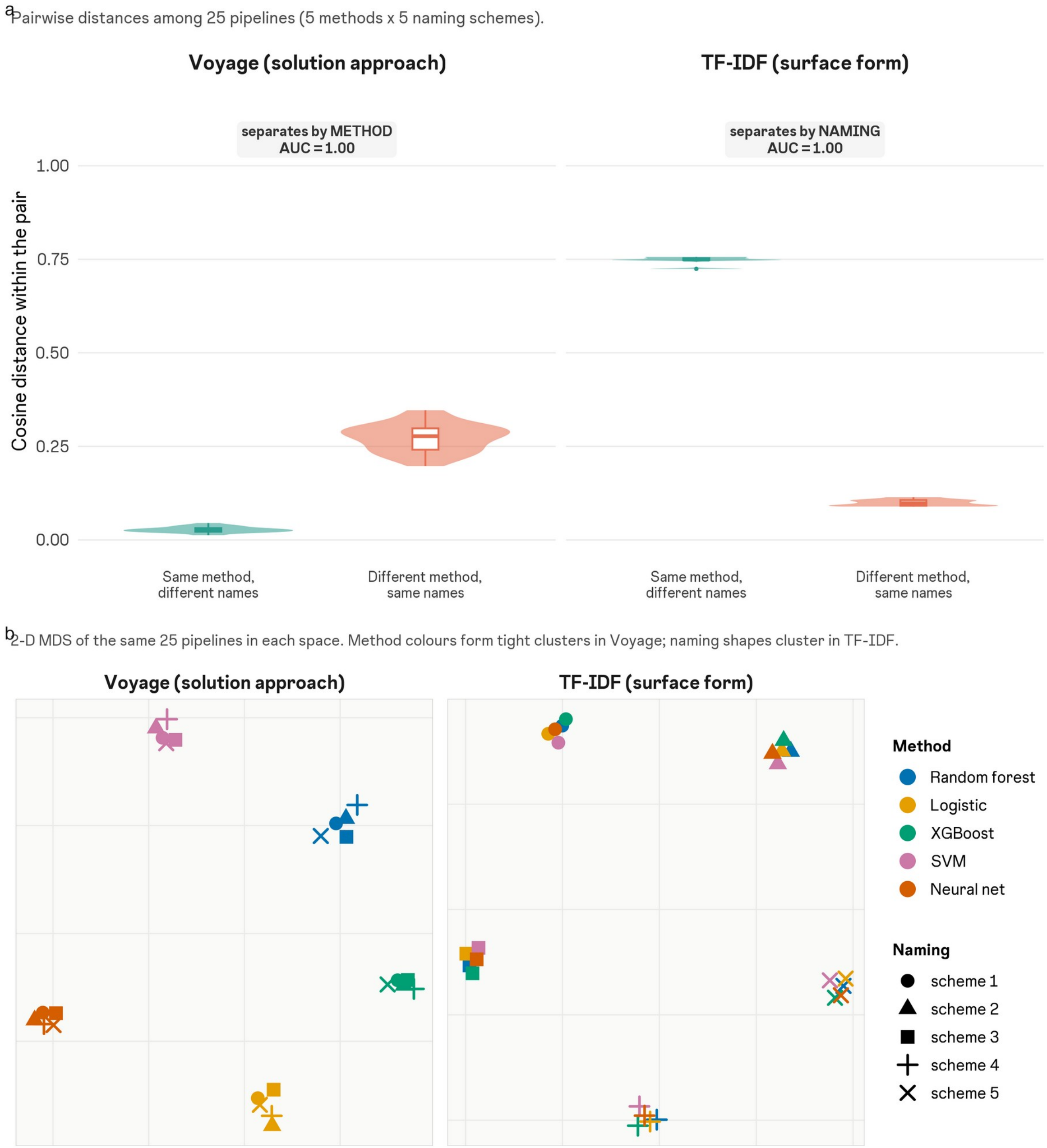

***Fig. S3. Embedding-measure validation on realistic machine-learning pipelines. a**, Distribution of pairwise distances for the 25 synthetic pipelines (five methods × five naming conventions), by pair type. **b**, Two-dimensional projection of the same pipelines in each space. Voyage separates by modeling approach; TF-IDF separates by naming convention (AUROC 1.00 in both cases).*

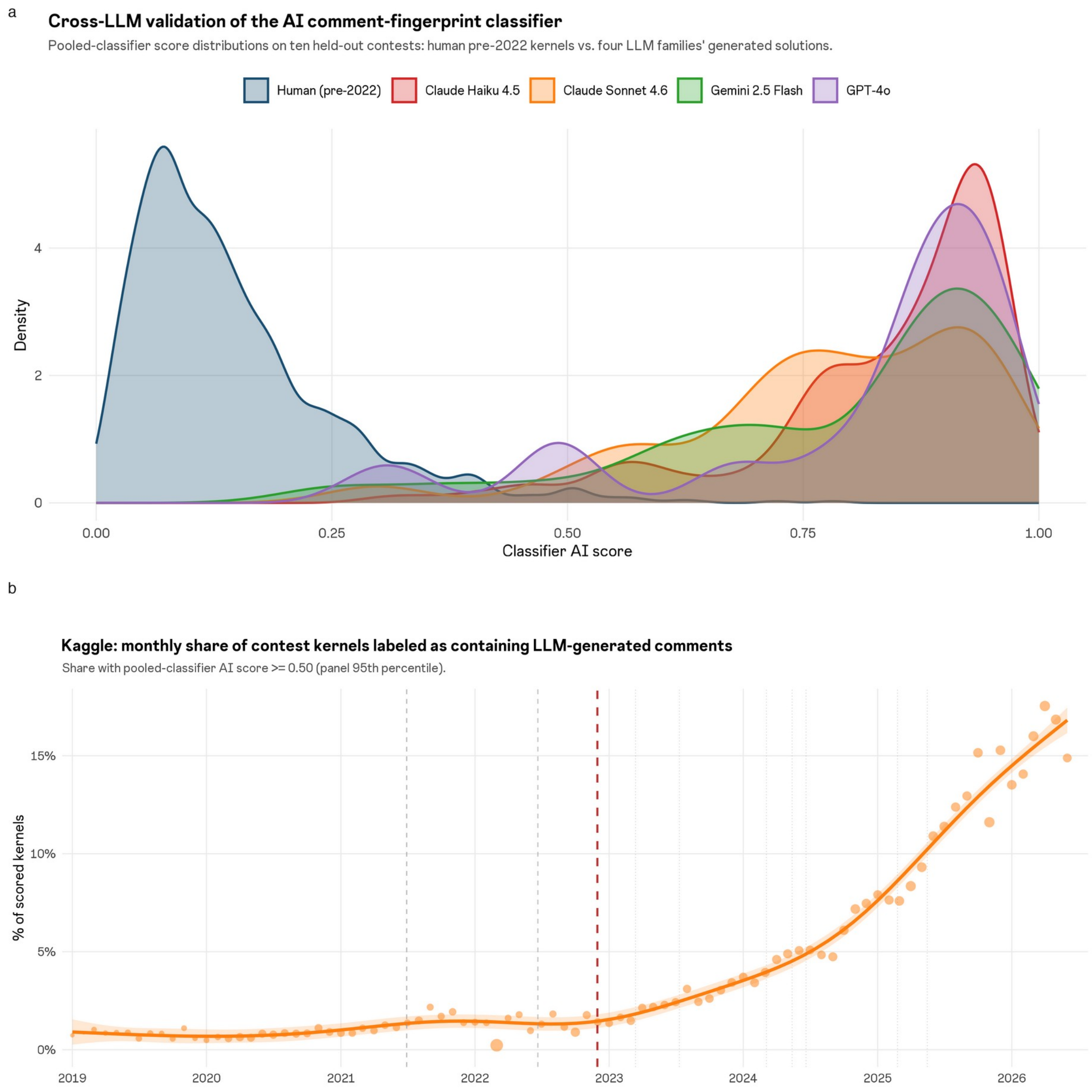

***Fig. S4. The comment classifier. a**, Score distributions for pre-2022 human kernels and for matched solutions from four language models, on ten held-out competitions. **b**, Share of scored kernels labeled as containing model-generated comments over time, rising from about 1% to about 17% by mid-2026. The post-release trend change is +0.37 percentage points per month ($P < 0.001$).*

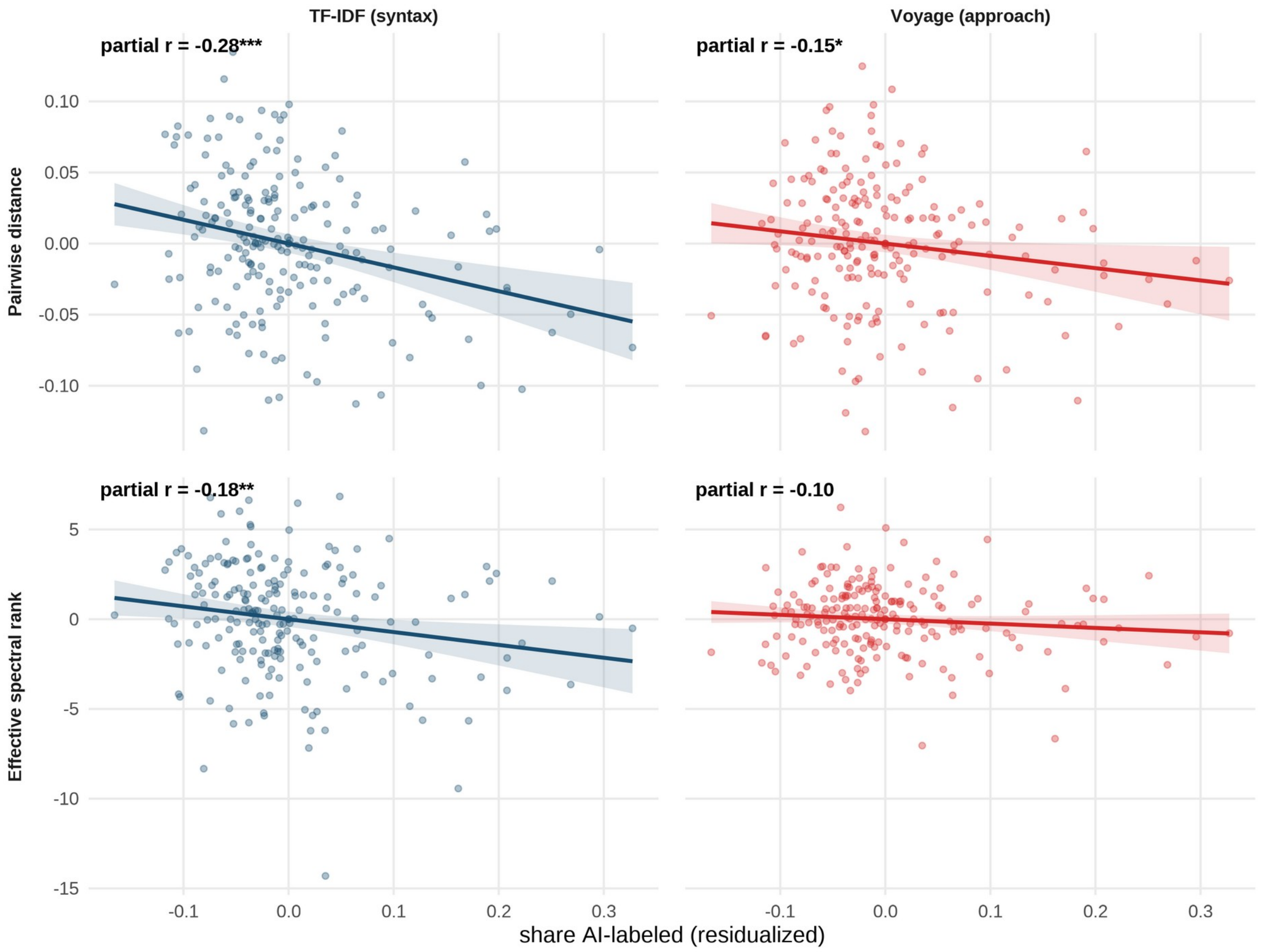

***Fig. S5. Contest-level association between measured AI usage and homogenization.*** *Each point is one competition (n = 205), residualized on the contest-description embedding and structured contest features. A ten-percentage-point increase in the share of labeled kernels is associated with a 0.017 reduction in TF-IDF pairwise distance (P < 0.001) and a 0.7-point reduction in TF-IDF effective rank (P < 0.01); the Voyage associations are weaker.*

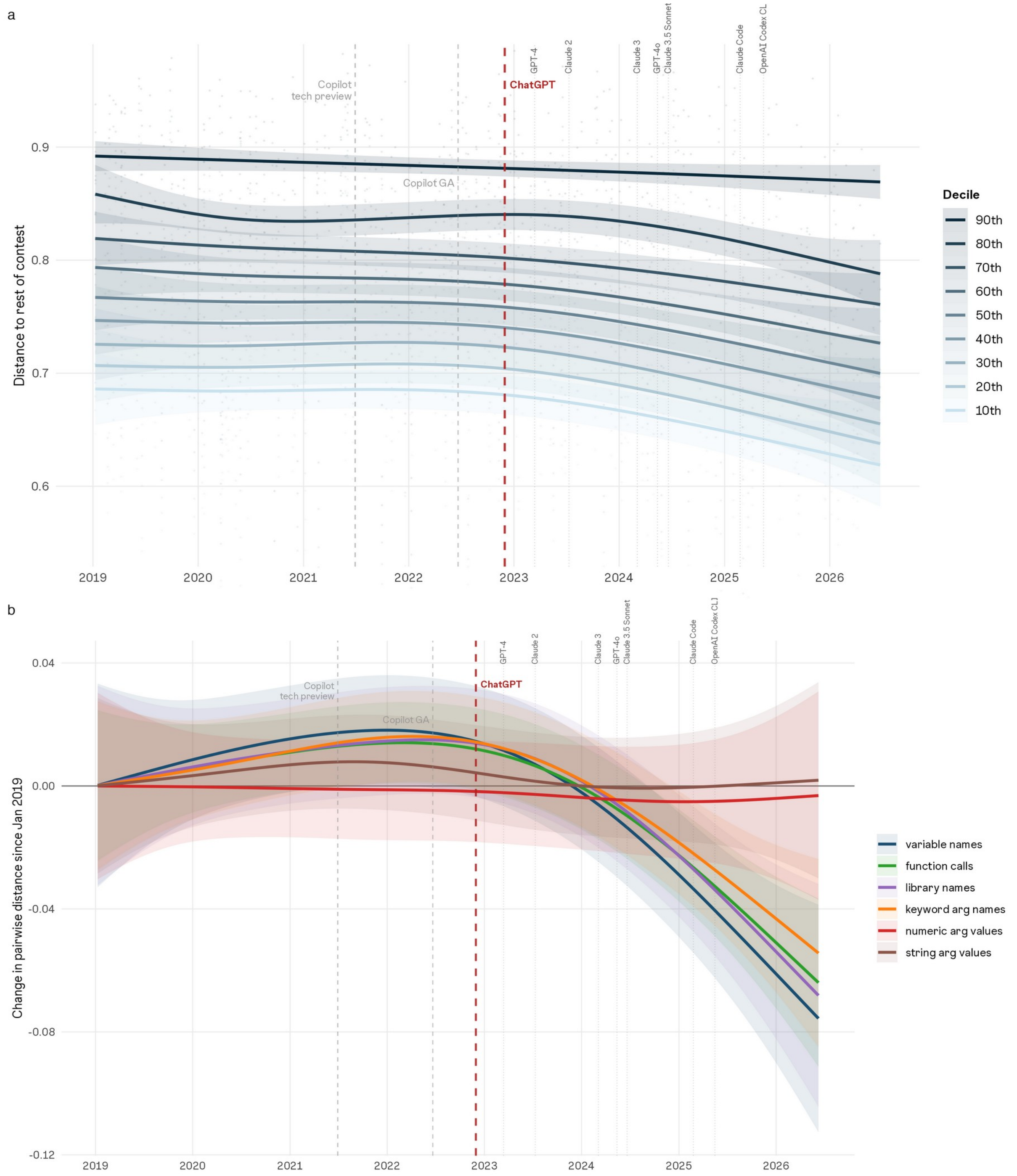

***Fig. S6. Decomposition of syntactic homogenization. a**, Within-contest TF-IDF distance by decile of the submission distribution ($n = 252$ competitions); convergence concentrates among already-conventional submissions. **b**, Convergence by code component, each measured in the original corpus's TF-IDF feature space; variable names, libraries, and function call components converge most (post-release trends $-0.0024$ to $-0.0031$, all $P < 0.001$).*

***Table S1. Embedding-measure validation by pair type.*** *Mean pairwise distance among the 25 synthetic pipelines.*

| **Pair type** | **Voyage (semantics)** | **TF-IDF (syntax)** |
|---|---|---|
| Same approach, different naming | 0.03 | 0.75 |
| Different approach, same naming | 0.27 | 0.10 |

***Table S2. Kernel-level association between LLM-comment score and seed-42 use.*** *Linear-probability estimates. Each cell reports the coefficient, its 95% confidence interval in brackets, and an exact p-value in parentheses, from standard errors clustered by competition. Columns are nested specifications on the same sample. The classifier-link analyses use a broader scored-kernel sample than the embedding analyses in the main text: 63,181 kernels from 458 competitions have valid AI-comment scores, and 27,732 of these kernels set at least one seed. The 61,565-kernel embedding sample instead applies the completed-contest, timing, and embedding-availability restrictions described above.*

| | (1) | (2) | (3) | (4) |
|---|---|---|---|---|
| AI-comment score | 0.308<br>[0.210, 0.406]<br>(P < 0.001) | 0.200<br>[0.125, 0.275]<br>(P < 0.001) | 0.125<br>[0.050, 0.199]<br>(P = 0.001) | 0.102<br>[0.031, 0.174]<br>(P = 0.005) |
| Author | Yes | Yes | Yes | Yes |
| Competition | No | No | Yes | Yes |
| Month | No | No | No | Yes |
| Sample | All scored | Seeded | Seeded | Seeded |
| Observations | 63,181 | 27,732 | 27,732 | 27,732 |

***Table S3. Robustness of the syntactic and semantic results to the embedding representation.*** *Post-release trend change per month. Each cell reports the estimate, its 95% confidence interval in brackets, and an exact p-value in parentheses, from heteroskedasticity-robust (HC3) standard errors. n = 252 (pairwise), 205 (effective rank).*

| **Representation** | **Type** | **Pairwise distance** | **Effective rank** |
|---|---|---|---|
| TF-IDF, token 1–2 gram | syntactic (main) | −0.0017 [−0.0029, −0.0005] (P = 0.005) | −0.160 [−0.239, −0.081] (P < 0.001) |
| TF-IDF, character 3 –5 gram | syntactic | −0.0032 [−0.0048, −0.0016] (P < 0.001) | −0.153 [−0.213, −0.093] (P < 0.001) |
| Cohere Embed v4 | semantic | 0.0003 [−0.0011, 0.0018] (P = 0.639) | 0.015 [−0.042, 0.071] (P = 0.610) |
| Voyage code-3 | semantic (main) | −0.0007 [−0.0021, 0.0006] (P = 0.284) | 0.044 [−0.019, 0.106] (P = 0.170) |

***Table S4. Convergence by code component.*** *Each component is measured relative to the original corpus's TF-IDF representation space (244 competitions, comment-stripped submissions). Exact p-values are in parentheses; the post-release trend also includes 95% confidence intervals in brackets. All tests use heteroskedasticity-robust (HC3) robust standard errors.*

| **Code component** | **Level, Jan 2019** | **Pre-release slope** | **Change in slope at release** | **Post-release trend** |
|---|---|---|---|---|
| Variable names | 0.840 | 0.0005<br>(P = 0.404) | −0.0034<br>(P < 0.001) | −0.0030<br>[−0.0041, −0.0018]<br>(P < 0.001) |
| Function calls | 0.868 | 0.0003<br>(P = 0.527) | −0.0029<br>(P < 0.001) | −0.0026<br>[−0.0036, −0.0016]<br>(P < 0.001) |
| Library names | 0.846 | 0.0001<br>(P = 0.838) | −0.0032<br>(P < 0.001) | −0.0031<br>[−0.0044, −0.0018]<br>(P < 0.001) |
| Keyword-argument names | 0.886 | 0.0004<br>(P = 0.327) | −0.0028<br>(P < 0.001) | −0.0024<br>[−0.0036, −0.0013]<br>(P < 0.001) |
| Numeric argument values | 0.930 | −0.0001<br>(P = 0.904) | −0.0002<br>(P = 0.804) | −0.0002<br>[−0.0012, 0.0007]<br>(P = 0.620) |
| String argument values | 0.837 | 0.0003<br>(P = 0.573) | −0.0001<br>(P = 0.854) | 0.0002<br>[−0.0009, 0.0012]<br>(P = 0.763) |

## Additional Analyses

In this SI Appendix, I report additional robustness checks addressing the association between within-contest homogeneity and collective overfitting, handling of near-duplicate Kaggle submissions, Kaggle submission forking and publication rates, and handling of embedding anisotropy. In Table S3, I also analyze the sensitivity of the null result in semantic space to the choice of embedding model.

**Homogenization and collective overfitting.** The basic notion of monoculture is that a lack of diversity leads to common failure modes and, in turn, ecosystem-level fragility. In a machine-learning competition, a primary failure mode is overfitting: a solution that scores well on the public test set may fail to generalize to the private test set. By reconstructing the full leaderboard for every competition in my sample from the Meta-Kaggle Teams and Submissions tables, I can relate within-contest code convergence to the observed drop in submissions' performance between the public and private leaderboards (Fig. S7 and Table S5). Competitions with more homogeneous submissions exhibit larger drops, and the relationship is concentrated in the upper tail rather than in typical competitions. Quantile regressions of the public-private drop on code homogeneity indicate coefficients of +0.21 at the 80th percentile ($P < 0.001$), +0.23 at the 90th ($P = 0.002$), and +0.30 at the 95th ($P = 0.05$), versus +0.09 for the mean ($P = 0.01$) and no significant relationship at the median.

Severe collective drops are roughly twice as common among the most homogeneous competitions, while the typical homogeneous competition is no worse than the typical diverse one. I obtain the same pattern whether I measure the drop at the competition or the team level, whether it is defined in signed or absolute terms, and after conditioning on competition size, prize, year, the contest-description embedding, the dispersion of public scores, and the breadth of approaches explored. In every case, I compute the public-private drop only using teams that did not post a public kernel. This ensures that no submission contributes to both the code homogeneity measure *and* the outcome measure. Considered in isolation, semantic homogeneity predicts a performance drop, on average (+0.099, $P = 0.03$), as does syntactic homogeneity (+0.077, $P = 0.06$); considered in tandem, both effects become statistically insignificant at conventional thresholds, though the relationship with semantic homogeneity is marginal (+0.080, $P = 0.11$ for semantic homogeneity versus +0.032, $P = 0.45$ for syntactic). Finally, a combined index of the two homogeneity measures (averaging them) is highly predictive of a performance drop, with a coefficient of +0.101 ($P = 0.02$).

My estimates suggest that semantic convergence is the greater source of fragility, though the two measures are highly correlated ($r = 0.63$) across competitions and the coefficients are not statistically distinguishable (difference +0.048, $P = 0.58$). Note that most of the variation in code homogeneity across competitions is dominated by cross-sectional differences in contest tasks and objectives, rather than the inter-temporal variation that would be attributable to AI's effects. I therefore cannot conclusively separate the effects of each homogeneity type with this sample. Within a given year, the two measures correlate at 0.72. Variation over time accounts for a much smaller share of the total: 7% of the variance in syntactic distance and 3% of the variance in semantic distance is intertemporal.

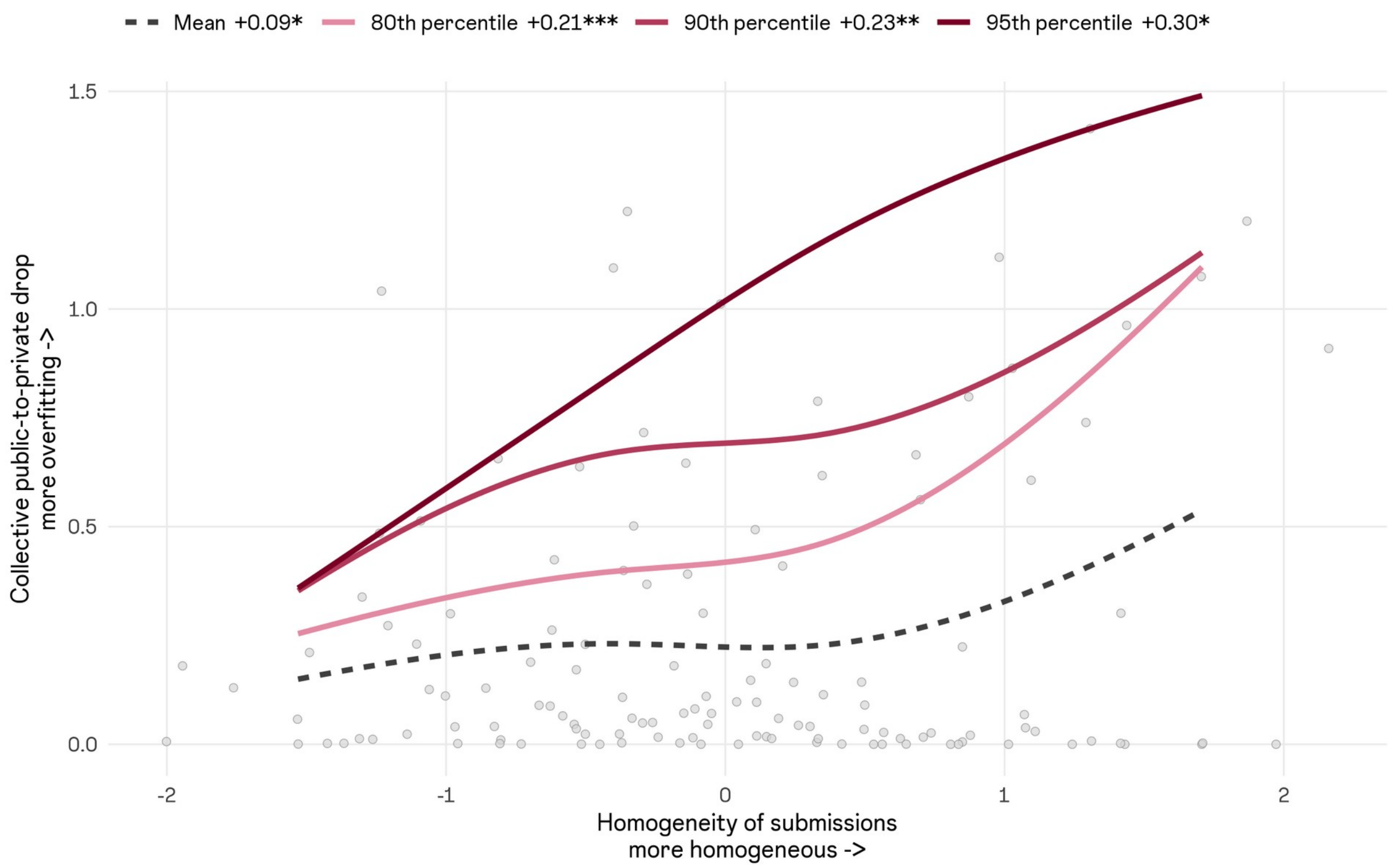

***Fig. S7. Percentiles of the collective public-to-private drop as a function of within-contest homogeneity.*** *One point per competition ($n = 129$). Homogeneity on the x-axis is the average of standardized within-contest syntactic and semantic pairwise distances, multiplied by -1 (negated), such that higher values reflect more homogeneous submissions; higher values on the y-axis imply that competitors overfit the public leaderboard and dropped when evaluated against the private holdout. The drop is computed only based on teams that posted no public kernel. As such, no submission contributes to both the homogeneity measure and the outcome calculation, ensuring the code homogeneity measure proxies for the submissions being evaluated for overfitting. Curves are quantile regressions drawn on a spline basis for readability; coefficients noted beside each series are based on linear fits (+ $P < 0.10$, * $P < 0.05$, ** $P < 0.01$, *** $P < 0.001$).*

***Table S5. Within-contest homogeneity and overfitting.*** *Each column reflects coefficients from a separate OLS regression of a competition's collective public-to-private performance drop on within-contest homogeneity; column (3) includes both homogeneity measures, jointly. Homogeneity is the reverse of standardized mean pairwise cosine distance among a competition's public submissions; a positive coefficient indicates that more homogeneous competitions suffer a larger drop. Every specification conditions on the number of teams, the prize, the competition year, and the principal components of the contest-description embedding. The drop is computed only over teams that posted no public kernel, i.e., no submission contributes to both the homogeneity measure and the outcome. The combined index averages the two standardized measures.*

| | (1) | (2) | (3) | (4) |
|---|---|---|---|---|
| Semantic homogeneity (Voyage dist) | 0.099<br>[0.011, 0.187]<br>(P = 0.028) | | 0.080<br>[−0.018, 0.179]<br>(P = 0.110) | |
| Syntactic homogeneity (TF-IDF dist) | | 0.077<br>[−0.004, 0.157]<br>(P = 0.063) | 0.032<br>[−0.052, 0.116]<br>(P = 0.450) | |
| Combined index (Mean of TF-IDF and Voyage dist) | | | | 0.101<br>[0.016, 0.186]<br>(P = 0.020) |
| Competitions | 129 | 129 | 129 | 129 |

**Near-duplicate submissions.** 'Near-duplicates' refers to distinct authors submitting nearly identical code (Jaccard $\geq 0.9$ when submissions are tokenized based on five-token shingles, i.e., where submissions are encoded based on a five-token sliding window over the entire submission content). Tokens are extracted using regular expression-based parsing; they capture whole words, individual variable or function names, and parameters. I find that near-duplicate submissions are uncommon, accounting for just 5.1% of all submissions, and that their prevalence falls post-AI. I obtain essentially the same within-contest syntactic convergence after removing them.

**Forking rate.** My interrupted time-series analysis of per-contest fork rates ($n = 253$ competitions; Fig. S8) indicates that fork rates were rising (+0.22 percentage points per month, $P = 0.004$) and that trend then reversed (post-release trend change of −0.26 percentage points per month, $P = 0.065$), with no significant level shift at the point of ChatGPT release itself ($P = 0.99$). I therefore find no evidence that rising fork rates explain post-release declines in average code novelty.

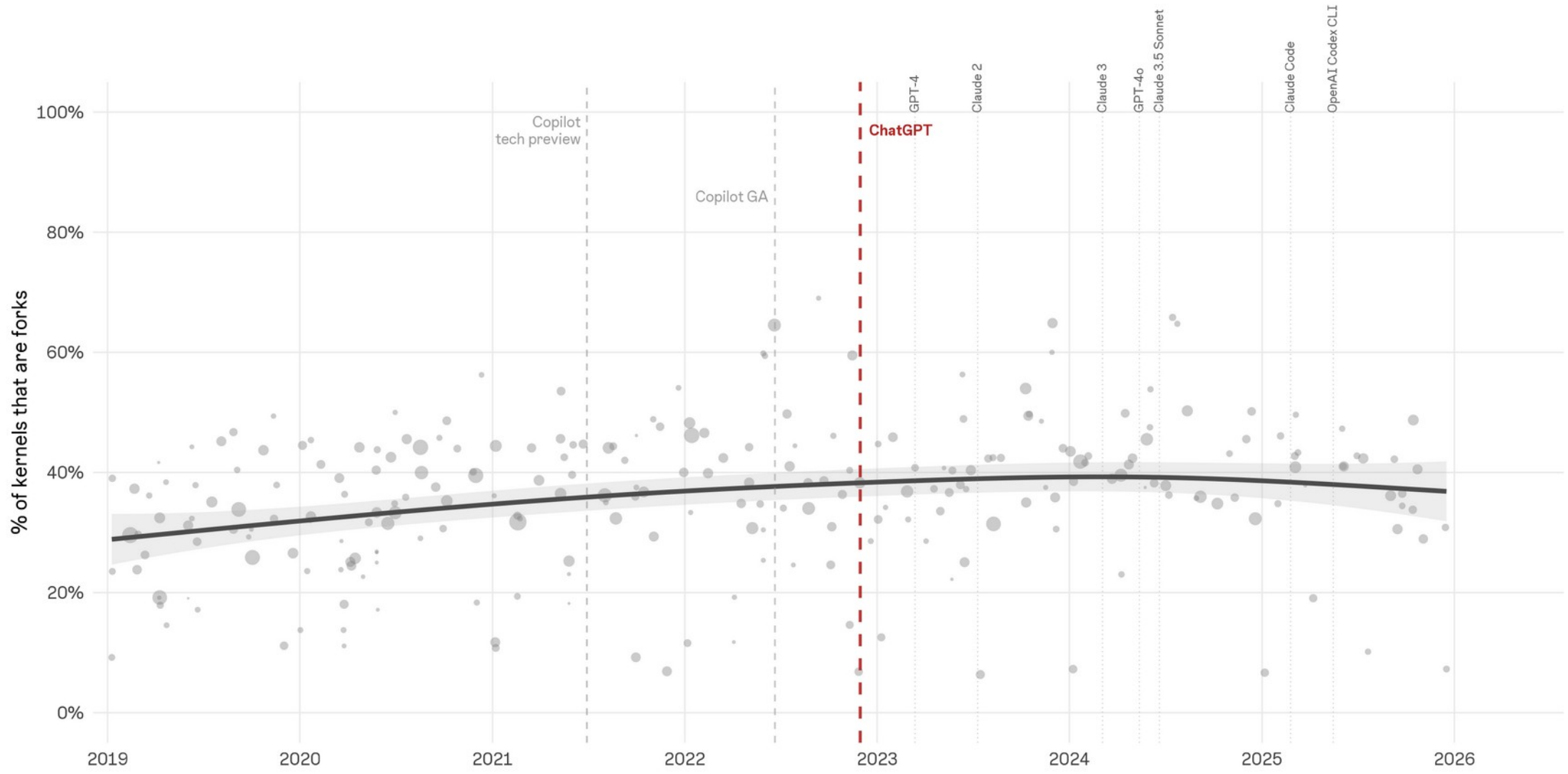

***Fig. S8. Forking rate per competition over time.*** *The percentage of Kaggle kernel submissions per contest that are explicit forks of other kernels (n = 253 competitions). The fraction rose through 2023, then began to decline. Forking does not appear to explain my findings of recent declines in code novelty.*

**Publication rate.** I find no significant change in per-contest publication rates (n = 205 competitions; Fig. S9) around the release of ChatGPT. The estimated level shift is insignificant (P = 0.89), as is the estimated trend change (P = 0.18). I therefore find it unlikely that shifts in publication behavior explain my findings.

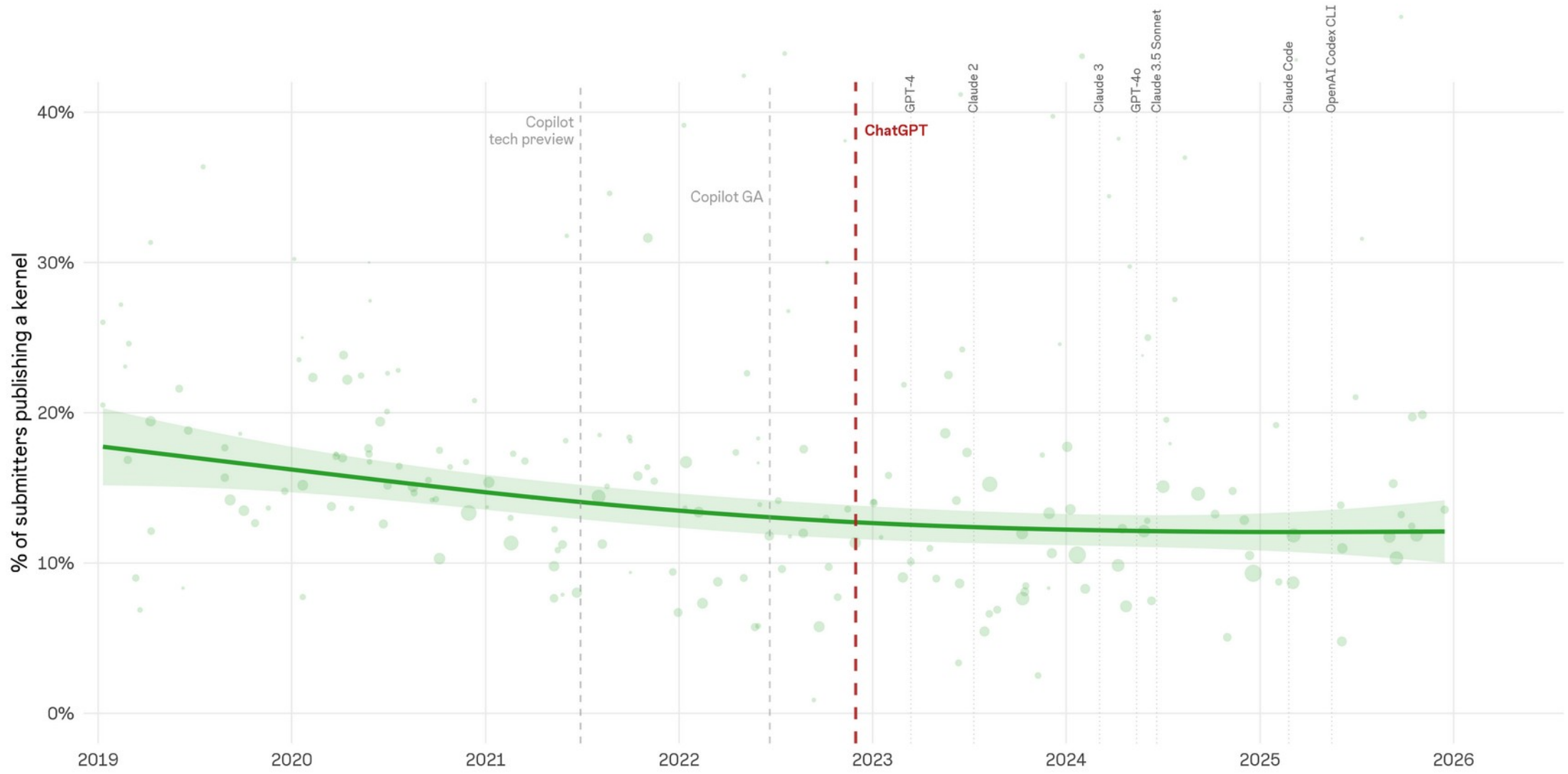

***Fig. S9. Publication rate per competition over time.*** *The percentage of Kaggle contest submitters, per contest, who publish at least one kernel (n = 205 competitions). The fraction declined until mid-2022, just before the emergence of ChatGPT, after which it has remained relatively flat, so shifts in Kaggle publication activity cannot explain my findings.*

**Embedding anisotropy and projection checks.** I find no indication that the null result in semantic (Voyage embedding) space is an artifact of embedding compression. The Voyage space's cross-contest standard deviation of pairwise distance (0.074) is nearly identical to TF-IDF's (0.075) and, within contest, distances span nearly the full cosine range (1st-to-99th percentile of 0.14 to 1.06). Expressed in standard deviations of each measure, the post-release change reflects 0.99 standard deviations in TF-IDF pairwise distance versus 0.43 in Voyage distance (1.78 versus 0.71 for effective rank). Thus, any semantic convergence is well under half syntactic convergence and is indistinguishable from zero.

I also find that the null result is robust to addressing possible embedding anisotropy. Contextual embeddings often occupy a narrow cone dominated by a few shared directions, compressing cosine distances. Voyage exhibits this common-direction anisotropy in raw form (the mean cosine distance between random submission pairs is just 0.47). The mean-centering I apply to all embedding measures removes the global common-mode component, which is the dominant source of distance 'deflation.' After centering, no single dimension dominates the embeddings; the top principal component explains just 6.4% of variance. Centering attenuates residual directional anisotropy rather than removing it entirely, however, so I also verify robustness directly. I find that projecting out the top one, two, and three global principal components has no impact on my conclusions about the lack of shifts in Voyage embeddings' distances or spectral rank; the pairwise-distance change remains within $\pm 0.0006$ per month, and the effective-rank change weakens toward zero. At the same time, the observed declines in TF-IDF cosine distance and effective rank remain when applying the same anisotropy-removal procedure.